\documentclass[twocolumn,showpacs,preprintnumbers,amsmath,amssymb]{revtex4}
\usepackage{graphicx}
\usepackage{dcolumn}
\usepackage{bm}

\newcommand{\bef}{\begin{figure}}
\newcommand{\eef}{\end{figure}}

\newcommand{\be}{\begin{equation}}
\newcommand{\ee}{\end{equation}}
\newcommand{\bea}{\begin{eqnarray}}
\newcommand{\eea}{\end{eqnarray}}
\usepackage{xspace}     
\usepackage{color}

\newcommand{\sqrts}[1]{\mbox{$\sqrt{s_{_{NN}}}$}~=~#1~GeV\xspace} 
\newcommand{\pp}{\mbox{$p$~+~$p$}\xspace} 
\newcommand{\AuAu}{\mbox{Au~+~Au}\xspace} 
\newcommand{\lra}{\leftrightarrow}

\begin{document}

\title{
Elliptic flow of $\phi$ meson a sensitive probe for onset of de-confinement transition in high energy heavy-ion collisions}

\author{Md. Nasim$^1$, Bedangadas Mohanty$^2$ and Nu Xu$^3$$^4$}
\affiliation{$^1$Variable Energy Cyclotron Centre, Kolkata 700064, India,
$^2$School of Physical Sciences, National Institute of Science Education and Research, Bhubaneswar 751005, India,
$^3$College of Physical Science and Technology, Central China Normal University, Wuhan 430079, China, and 
$^4$Nuclear Science Division, Lawrence Berkeley National Laboratory, Berkeley, CA 94720, USA}

\date{\today}
\begin{abstract}

Elliptic flow  ($v_2$) of $\phi$ meson is shown to be a sensitive probe of the
partonic collectivity using A Multi Phase Transport (AMPT) model. Within the
ambit of the AMPT model with partonic interactions (string melting version),
the $\phi$ meson $v_2$ at midrapidity is found to have negligible contribution from hadronic
interactions for \AuAu collisions at \sqrts{200}. Changing the hadron cascade time in
the model calculations does not change the $\phi$ meson $v_2$, while it reduces
the $v_2$ of proton at low transverse momentum ($p_{T}$).
These observations indicate that a substantial reduction in  $\phi$ mesons $v_2$ as
a function of colliding beam energy would suggest the dominance of hadronic interactions over
partonic interactions.

\end{abstract}
\maketitle

\section{Introduction}
\label{sec:introduction}

One of the current focus of the high energy heavy-ion collision experiments
is to study the various aspects of the QCD phase diagram~\cite{Gupta:2011wh,Mohanty:2011ki}.
After observing the clear signatures of the formation of strongly interacting
quark-gluon plasma (QGP) at the top RHIC energies~\cite{Arsene:2004fa,Back:2004je,Adams:2005dq,Adcox:2004mh},
attempts are being made to vary the colliding beam energy and
search for the transition region between the partonic and/or hadronic dominant interactions
in the phase diagram. This is one of the goals of the RHIC Beam Energy Scan (BES)
program~\cite{starnote,Mohanty:2011nm}.

In \AuAu collisions at \sqrts{200} the $\phi$ meson production has played a
crucial role to establish the formation of the partonic matter~\cite{Mohanty:2009tz}.
Number of constituent quark scaling of the elliptic flow of $\phi$ meson~\cite{Abelev:2007rw},
enhancement in yield of $\phi$ meson in \AuAu collisions relative to \pp collisions~\cite{Abelev:2008zk}
and the ratio of yield of the $\Omega$ baryon to the yield of $\phi$ meson
as a function of $p_{T}$~\cite{Abelev:2007rw} have been the key measurements.
In general, multi-strange hadrons are understood~\cite{vanHecke:1998yu} to have a smaller hadronic interaction
cross sections. As a result, their momentum distributions, and particularly the momentum
dependence of the azimuthal anisotropy~\cite{Abelev:2007rw}, are expected to be primarily controlled by the early
partonic interactions in high-energy nuclear collisions.
In order to demonstrate that the multi-strange hadrons, are a clean tool for phase identification,
we have performed model simulations to study the $\phi$ meson production.
The paper focuses on using the AMPT model for the $\phi$ meson production, leading to the establishment
of the $\phi$ $v_2$  as a key observable for studying the onset of the de-confinement transition.

Further, with high statistics data being collected in high energy heavy-ion collisions, it is now
possible to have high precision measurements of $v_2$ for various produced hadrons.
One possibility that opens up is to study the effect of the late stage hadronic re-scattering
on $v_2$ at low $p_{T}$. Initial simulations using a hybrid model showed
that the usual mass ordering trend of $v_{2} (\phi)$ $<$ $v_{2} (p)$ will get
reversed due to the late stage hadronic re-scattering~\cite{Hirano:2007ei}.
In this paper we also study this aspect using the AMPT model.

  The paper is organized as follows.
  In the section~\ref{sec:ampt}, we will briefly introduce the AMPT model~\cite{Lin:2004en}  used in this study.
  In the section~\ref{sec:results}, the results from the AMPT model calculations
  regarding $v_{2}$ of $\phi$ mesons and protons for various configurations
  (different parton-parton cross section and hadronic cascade time) are presented.
  Finally in section~\ref{sec:summary}, we summarize our findings and present a
  short discussion on the implications of this work to the current experimental
  measurements in high energy heavy-ion collisions.

\section{Model Calculations}
\label{sec:ampt}

\subsection{AMPT}

The AMPT~\cite{Lin:2004en} model  used for the calculations presented in this paper has
four main stages: the initial conditions, partonic interactions, the conversion
from the partonic to the hadronic matter, and hadronic interactions.
The initial conditions are obtained from the HIJING model~\cite{Gyulassy:1994ew}.
Scatterings among partons are modeled by Zhang's parton cascade
(ZPC)~\cite{Zhang:1997ej}, it includes only two-body scatterings with cross sections
obtained from the pQCD with screening masses. Some of the results presented
are by varying the parton-parton scattering cross sections within 3 mb to 14 mb.
The AMPT model with string melting~\cite{Lin:2001zk} leads to hadron formation using a quark coalescence model.
The subsequent hadronic matter interaction is described by a hadronic cascade,
which is based on A Relativistic Transport (ART) model~\cite{Li:1995pra}. The termination time of the
hadronic cascade is varied in this paper from 0.6 fm/$c$ to 30 fm/$c$ to study the
effect of the hadronic re-scattering on the observables presented.
More detailed discussions regarding the AMPT model can be found in Ref.~\cite{Lin:2004en}.
In this study, approximately one million events for each configuration (different cross section and
hadronic cascade time) were generated for \AuAu 0-80\% minimum bias collisions
at \sqrts{200}. All results presented are for the rapidity range $\pm$ 1.0 unit.

\subsection{$\phi$ mesons production}

The string melting version of the AMPT used in this paper,
produces $\phi$ meson using a quark coalescence model in the
partonic stage~\cite{Lin:2004en}.  In the hadronic stage, the AMPT model includes
the following reactions associated with
the $\phi$ meson. Inelastic scatterings in baryon-baryon channels
includes $(N \Delta N^*) (N \Delta N^*) \rightarrow \phi NN$,
those in the meson-baryon channel includes
$(\pi \rho) (N \Delta N^*) \lra \phi (N \Delta N^*)$ and
$K (\Lambda \Sigma) \lra \phi N$. The $\phi$ meson scatterings
with other hadrons included in the model are
$\phi (\pi \rho \omega) \lra (K K^*)(\bar K \bar {K^*})$, and
$\phi (K K^*) \lra (\pi \rho \omega) (K K^*)$.
The cross section for the elastic
scattering of the $\phi$ meson with a nucleon is set to 8 mb
while the $\phi$ meson elastic cross section with a meson is set to 5 mb.
For other details and specifically those related to inelastic
scattering cross section can be found in~\cite{Lin:2004en}.

For comparison with another transport based calculation, we also present $v_{2}$ for 
$\phi$ meson from UrQMD (Ultra relativistic Quantum Molecular Dynamics) model~\cite{urqmd}
for Au+Au collisions at $\sqrt{s_{\mathrm {NN}}}$ = 200 GeV. 
It is based on a microscopic transport theory where the phase space description of the
reactions are important. It allows for the propagation of all hadrons on classical 
trajectories in combination with stochastic binary scattering, color string
formation and resonance decay. It incorporates baryon-baryon, meson-baryon and meson-meson 
interactions, the collisional term includes more than 50 baryon species and 45 meson species.

\subsection{Elliptic flow }

The anisotropic elliptic flow parameter presented in this paper is defined as
the $2^{\mathrm {nd}}$ Fourier coefficient $v_{2}$ of the particle distributions
in emission azimuthal angle ($\phi$) with respect to the reaction plane angle
($\Psi$)~\cite{Voloshin:1994mz}, and can be written as
\begin{equation} \frac{dN}{d\phi} \propto
1+2 v_2\cos(2(\phi - \Psi)). \end{equation}
For a given rapidity window the second coefficient is
\begin{equation}
v_{2}=\langle\cos(2(\phi-\Psi))\rangle=\langle\frac{p_x^2-p_y^2}{p_x^2+p_y^2}\rangle,
\end{equation}
where $p_x$ and $p_y$ are the $x$ and $y$ component of the particle momenta. In the AMPT
model the $\Psi$ is along the x-axis.

\section{Results}
\label{sec:results}

  \begin{figure}[htbp]
\includegraphics[scale=0.4]{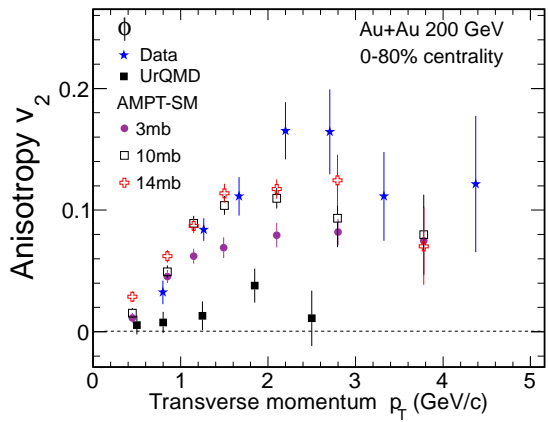}
    \caption{\label{fig:fig1}
      (Color online) $\phi$ meson $v_{2}$ for \AuAu minimum bias (0-80\%) collisions at midrapidity ($\pm$ 1.0)
     at \sqrts{200} from STAR experiment at RHIC~\cite{Abelev:2007rw} compared to the corresponding AMPT and UrQMD model calculations. The
     AMPT model calculations are shown for three different parton-parton interaction cross sections
     of 3, 10 and 14 mb. The errors shown are statistical.
     }
  \end{figure}
    Figure~\ref{fig:fig1} shows the comparison of elliptic flow of $\phi$ meson in 0-80\% minimum bias
   \AuAu collisions at midrapidity for \sqrts{200} measured by STAR experiment at RHIC~\cite{Abelev:2007rw}
   with the corresponding
   results from the AMPT model for three different values of parton-parton corss sections of 3 (magenta solid circle),
   10 (black open square), and 14 mb (red open cross).
   Although it is generally believed that perturbative QCD cross section is about 3 mb~\cite{Molnar:2001ux}, in order to
   explain the measurements for $p_{T}$ $>$ 1 GeV/$c$ a parton-parton cross section between 10-14 mb is
   found to be required. Thus the generation of this  substantial $v_{2}$ for $\phi$  mesons as observed in
   the experiments requires a significantly large parton interaction cross section than obtained from pQCD
   calculations ($\sim$ 3 mb)~\cite{Lin:2004en,Molnar:2001ux,Zhang:1999rs}. Also shown in the figure is the
   corresponding $v_{2}$ results from the UrQMD model~\cite{urqmd}. The UrQMD model results (which does not include
   any partonic interactions) gives substantially smaller value of $v_{2}$ compared to the experimental data.

   Within the framework of hydrodynamics, the
   collectivity reflected through the $v_{2}$ distributions are caused by the pressure gradients. Due to
   the effect of self quenching, one expects that the development of $v_{2}$ is dominantly from the early stages
   of the collision~\cite{Zhang:1999rs,Sorge:1998mk}. From the comparison shown in Fig. 1, we conclude
   that collectivity in the multi-strange hadron $\phi$ meson has been developed in the early
   partonic interactions in high-energy nuclear collisions at RHIC~\cite{Arsene:2004fa}.
   We now proceed to investigate the contributions to $\phi$ meson $v_{2}$
   in AMPT separately from partonic and hadronic interactions.

\subsection{$\phi$ meson $v_{2}$ from partonic interactions}

  \begin{figure}[htbp]
\includegraphics[scale=0.4]{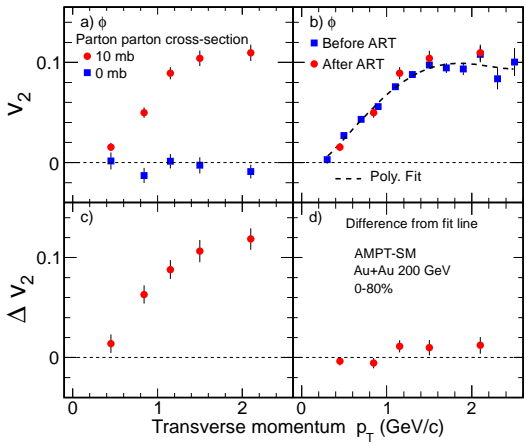}
    \caption{\label{fig:fig2}
      (Color online) $\phi$ meson $v_{2}$ for \AuAu minimum bias (0-80\%) collisions at midrapidity ($\pm$ 1.0)
     at \sqrts{200} from the AMPT model. Panels (a) and (b) shows the results as a function of $p_{T}$ for
     parton-parton interaction cross section of 0 and 10 mb and calculations before and after relativistic
     transport calculations for hadrons, respectively. The lower panels (c) and (d) shows the difference in $v_{2}$
     shown in panels (a) and (b), respectively. The errors shown are statistical.}
  \end{figure}

Figure~\ref{fig:fig2} (a) shows the $\phi$ meson $v_{2}$ for minimum bias \AuAu collisions at midrapidity
versus $p_{T}$ from AMPT model for parton-parton cross section
of 10 mb (red solid circles) and results without any parton-parton interaction (blue solid square, obtained
by setting the parton-parton cross section value to 0 mb). The hadronic cascade time is 30 fm/$c$ for both
the cases. The  $\phi$ meson $v_{2}$ is consistent with zero in absence of parton-parton interactions.
The panel (c) shows the difference between the two results, indicating that almost all the  $\phi$ meson $v_{2}$
is generated via the partonic interactions.

Figure~\ref{fig:fig2} (b) and (d) reinforces these observations by presenting the $\phi$ meson $v_{2}$ for
the same system before (solid blue squares) and after (solid red circles) the relativistic transport
calculations for hadrons. The results are similar between the two cases as seen from the
$p_{T}$  dependence of $\phi$ meson $v_{2}$ in panel (b), the difference between the two cases
shown in panel (d) is consistent with zero. These results indicate that partonic interactions
are essential for generating $v_{2}$ of $\phi$ mesons and contributions from hadronic interactions
are minimal,  within the context of the AMPT calculation.

  \subsection{Effect of hadronic re-scattering}
  \label{subsec:elliptic_flow}

  \begin{figure}[htbp]
\includegraphics[scale=0.4]{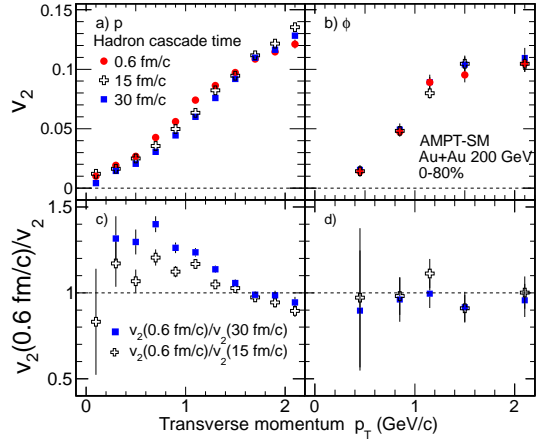}
    \caption{\label{fig:fig3}
      (Color online) (a) $v_{2}$ of protons as a function of $p_{T}$ for \AuAu 0-80\% collisions
      at \sqrts{200} from AMPT model at midrapidity. The results are shown for a parton-parton
     cross section of 10 mb and three different values of hadronic cascade time periods.
      (b) The same plot as (a) for the $\phi$ mesons.
      (c) Ratio of $v_{2}$ of protons for hadron cascade time of 0.6 fm/$c$ to corresponding  $v_{2}$
         for time periods of 15 and 30 fm/$c$,
      and (d) same as in (c) for the $\phi$ mesons. The error bars shown are statistical.
    }
  \end{figure}

To further study the effect of hadronic interactions, the model simulations were carried out for the
\AuAu minimum bias collisions with parton-parton interaction cross section fixed to be 10 mb and varying
the hadronic cascade time from 0.6 fm/$c$ to 30 fm/$c$. Higher value of hadronic cascade time reflects
larger hadronic re-scatterings. We have checked that for RHIC energies going to even longer time
duration does not contribute any further to the results presented. The $v_{2}$ calculations
are carried out for both $\phi$ meson and proton. We chose protons mainly for two reasons:
(a) as a hadron, it has a mass similar to that of the $\phi$ meson and
(b) contrary to that of the $\phi$ meson, it has larger hadronic interaction cross sections.

    Figure~\ref{fig:fig3} (a) shows the $v_{2}$ of protons versus $p_{T}$ for 0-80\% \AuAu collisions
    at \sqrts{200} from the AMPT model with parton-parton interaction cross section of 10 mb and
   three different values of hadronic cascade time of 0.6 fm/$c$ (red solid circle), 15 fm/$c$ (black
   open cross) and 30 fm/$c$ (blue solid square). With increase in hadron cascade time, which reflects
   increasing contributions from hadronic interactions, the proton $v_{2}$ decreases at lower $p_{T}$.
   Implying the development of the collective expansion in the hadronic period.
   This is more clearly illustrated in the panel (c) of the figure, which shows the ratio of the proton
   $v_{2}$ for the hadron cascade time of 0.6 fm/$c$ to the corresponding  $v_{2}$ values for time periods
   of 15 (open crosses) and 30 (solid squares) fm/$c$.  Figure~\ref{fig:fig3} (b) and (d) shows the corresponding results for
   $\phi$ mesons. In marked contrast to the case for protons, the $\phi$ meson $v_{2}$ remains unaffected
   by the hadronic interactions, indicating that $v_{2}$ is solely generated
   due to the partonic interactions in these model calculations.

  \begin{figure}[htbp]
\includegraphics[scale=0.4]{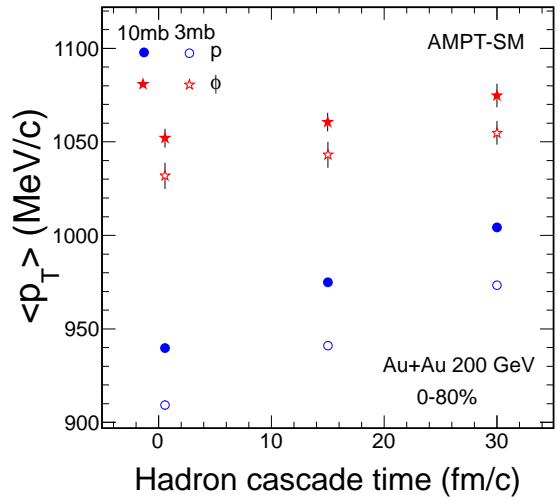}
    \caption{\label{fig:fig4}
      (Color online) Average transverse momentum ($< p_{T} >$) for proton and $\phi$ meson
      produced in 0-80\% \AuAu collisions at midrapidity for \sqrts{200} from AMPT model.
      The results are shown as a function of hadron cascade time and for two values of
      parton-parton cross sections.
    }
  \end{figure}
  Figure~\ref{fig:fig4} shows the effect of hadron re-scattering on the mean values of the
  transverse momentum distributions for proton and $\phi$ meson. The average transverse momentum
  $< p_{T} >$ for protons increases as the hadronic cascade time increases, where as
  for $\phi$ mesons this change is much smaller. The results are shown for two different
  values of parton-parton cross section values of 3 and 10 mb. For both cases,
  the trends are identical, suggesting that the increase in value of $< p_{T} >$
  with hadron cascade time is dominantly due to hadronic interactions. The magnitude of
  $< p_{T} >$ though is decided by both partonic and hadronic interactions taken
  together. Hence the $< p_{T} >$ for calculations with  parton-parton cross section
  of 3 mb is lower compared to those with parton-parton cross section of 10 mb.

\section{Summary}
\label{sec:summary}

  In summary, we have studied the generation of elliptic flow in AMPT model
  for $\phi$ mesons at midrapidity for 0-80\% centrality \AuAu collisions at
  \sqrts{200}. The large value of $v_{2}$ for  $\phi$ meson from the model
  which is comparable to the corresponding measurements at RHIC is due to
  the partonic interactions. Hadronic interactions as modelled in AMPT are unable to
  generate any $v_{2}$ for $\phi$ meson. Increasing the hadronic re-scattering by increasing the
  hadron cascade time period does not change the conclusions. In contrast to
   $\phi$ meson, the $v_{2}$ of protons decreases with increase in hadronic
   re-scattering effects at low transverse momentum. This effect is also
   seen in the average transverse momentum of protons which increases
   with increase in the time duration of the hadron cascade.

We have identified the multi-strange hadron $\phi$ meson as a clean tool
for studying partonic interactions in high-energy nuclear collisions.
Any significant change in $\phi$ meson collectivity will signal the possible
change of in the nature of the medium formed in high energy heavy-ion collisions.
  Zero value of the $v_{2}$ for the $\phi$ meson would clearly indicate that the system
  formed in heavy-ion collisions did not make the de-confinement transition.
  Hence $\phi$ meson $v_{2}$ is an ideal probe for the search of the phase boundary
  in the QCD phase diagram.

  In addition our study shows that proton $v_{2}$ at low $p_{T}$ decreases
  with increase in hadron cascade time. This reflectes the effect of late
  stage hadronic re-scattering. Whereas for the $\phi$ meson the $v_{2}$
  remains unaffected by late stage hadronic re-scattering process.
  In experiments it has been observed that at low $p_{T}$ a distinct
  mass ordering of $v_{2}$ is followed. Heavier particles have
  smaller $v_{2}$~\cite{Abelev:2008ae}. If re-scattering is sufficiently large the
  proton $v_{2}$ at low $p_{T}$  could be smaller than the corresponding
  $v_{2}$ for the $\phi$ meson. The key fact being that the $v_{2} (\phi)$ remains
unchanged with increase in hadronic re-scattering, while the $v_{2} (p)$ decreases
at low $p_{T}$.
  This can then be considered to be a signature
  of hadronic re-scattering and can be probed using high event statistics data sets
  in the high energy heavy-ion collision experiments.

\noindent{\bf Acknowledgments}\\
We thank Dr. Zi-Wei Lin for useful discussions on AMPT model results.
BM is supported by the DST SwarnaJayanti project fellowship.

\end{document}